\documentclass[lettersize, journal]{IEEEtran}
\IEEEoverridecommandlockouts
\usepackage{cite}
\usepackage{amsmath,amssymb,amsfonts}

\usepackage{flushend}

\usepackage[hidelinks,colorlinks=true,linkcolor=blue,citecolor=cyan]{hyperref}

\usepackage{verbatim}

\usepackage{mathtools, nccmath}

\usepackage{nccmath}

\usepackage{amsmath}
\usepackage{mathtools}

\usepackage{bbding}
\usepackage{pifont}

\usepackage{balance}

\usepackage{multirow}

\usepackage{ragged2e}

\usepackage{arydshln}
\usepackage{relsize}
\usepackage{hhline}
\usepackage{colortbl}

\usepackage[normalem]{ulem}

\usepackage{numprint}

\usepackage{stmaryrd}

\usepackage[hang,flushmargin]{footmisc} 

\usepackage{tikz}

\usepackage{algorithmic}
\usepackage{graphicx}
\usepackage{textcomp}
\usepackage{xcolor}
\def\BibTeX{{\rm B\kern-.05em{\sc i\kern-.025em b}\kern-.08em
    T\kern-.1667em\lower.7ex\hbox{E}\kern-.125emX}}

\usepackage{tikz,xcolor}

\definecolor{lime}{HTML}{A6CE39}
\DeclareRobustCommand{\orcidicon}{%
	\begin{tikzpicture}
	\draw[lime, fill=lime] (0,0) 
	circle [radius=0.16] 
	node[white] {{\fontfamily{qag}\selectfont \tiny ID}};
	\draw[white, fill=white] (-0.0625,0.095) 
	circle [radius=0.007];
	\end{tikzpicture}
	\hspace{-2mm}
}

\foreach \x in {A, ..., Z}{%
	\expandafter\xdef\csname orcid\x\endcsname{\noexpand\href{https://orcid.org/\csname orcidauthor\x\endcsname}{\noexpand\orcidicon}}
}

\newcommand{\Design}{\textbf{\texttt{VDC-2\textsuperscript{$n$}}}}

\begin{document}

\title{\huge 

Improved Data Encoding for Emerging Computing Paradigms: \\ From Stochastic to Hyperdimensional Computing

}


\author{
Mehran~Shoushtari~Moghadam\textsuperscript{\ding{69}} \orcidA{},~\IEEEmembership{Graduate Member,~IEEE,} Sercan~Aygun\textsuperscript{\ding{105}} \orcidB{},~\IEEEmembership{Senior Member,~IEEE,} and M.~Hassan~Najafi\textsuperscript{\ding{69}} \orcidE{},~\IEEEmembership{Senior Member,~IEEE}

\thanks{
This work was supported in part by National Science Foundation (NSF) grants \#2019511, \#2339701, 
and generous gift from 
Nvidia.
\\
\noindent 
\scriptsize{\ding{69}: Electrical, Computer, and Systems Engineering Department, Case Western Reserve University, Cleveland, OH 44106, USA. E-mail:\{moghadam, najafi\}@case.edu.\\ \ding{105}: School of Computing and Informatics, University of Louisiana at Lafayette, Lafayette, LA 70503, USA. E-mail:sercan.aygun@louisiana.edu.}
}
}

\maketitle

\begin{abstract}
    Data encoding is a fundamental step in emerging computing paradigms, 
    particularly in stochastic computing (SC) and hyperdimensional computing (HDC), where it plays a crucial role in determining the overall system performance and hardware cost efficiency. This study presents an advanced encoding strategy that leverages a 
    hardware-friendly class of low-discrepancy (LD) sequences, specifically 
    powers-of-2 bases of Van der Corput (VDC) sequences (\Design), as sources for random number generation. Our approach significantly enhances the accuracy and efficiency of SC and HDC systems by addressing challenges associated with randomness. By employing LD sequences, we improve correlation properties and reduce hardware complexity. Experimental results demonstrate significant improvements in accuracy and energy savings for SC and HDC systems. 
    Our solution provides a robust framework for integrating SC and HDC in resource-constrained environments, paving the way for efficient and scalable AI implementations.
\end{abstract}


\section{Introduction}
\IEEEPARstart{P}{roper} data encoding has always been a challenging task as the fundamental step in 
emerging computing models. 
Stochastic computing (SC) and hyperdimensional computing (HDC) have emerged as two promising 
paradigms for the efficient hardware design of 
machine learning systems. Both SC and HDC utilize long streams of `0's and `1's,  rather than conventional binary values (with positional encoding or bit-significance) 
as their basic computational elements.
Bit-streams ($\mathcal{BS}$s) and hypervectors ($\mathcal{HV}$s) serve
as the primitive components, acting as atomic data elements in SC and HDC, respectively. These atomic data elements 
are generated using a \textit{proper} source of randomness.
The state-of-the-art (SOTA) approaches typically employ pseudo-random number generators as 
the source of randomness to generate  
$\mathcal{BS}$s~\cite{Trig-Parhi,9444648} or $\mathcal{HV}$s~\cite{LeHDC, onlinehd, cascadehd,goktug-glsvlsi}. 

\begin{figure}[t]
  \centering
  \includegraphics[width=\linewidth]{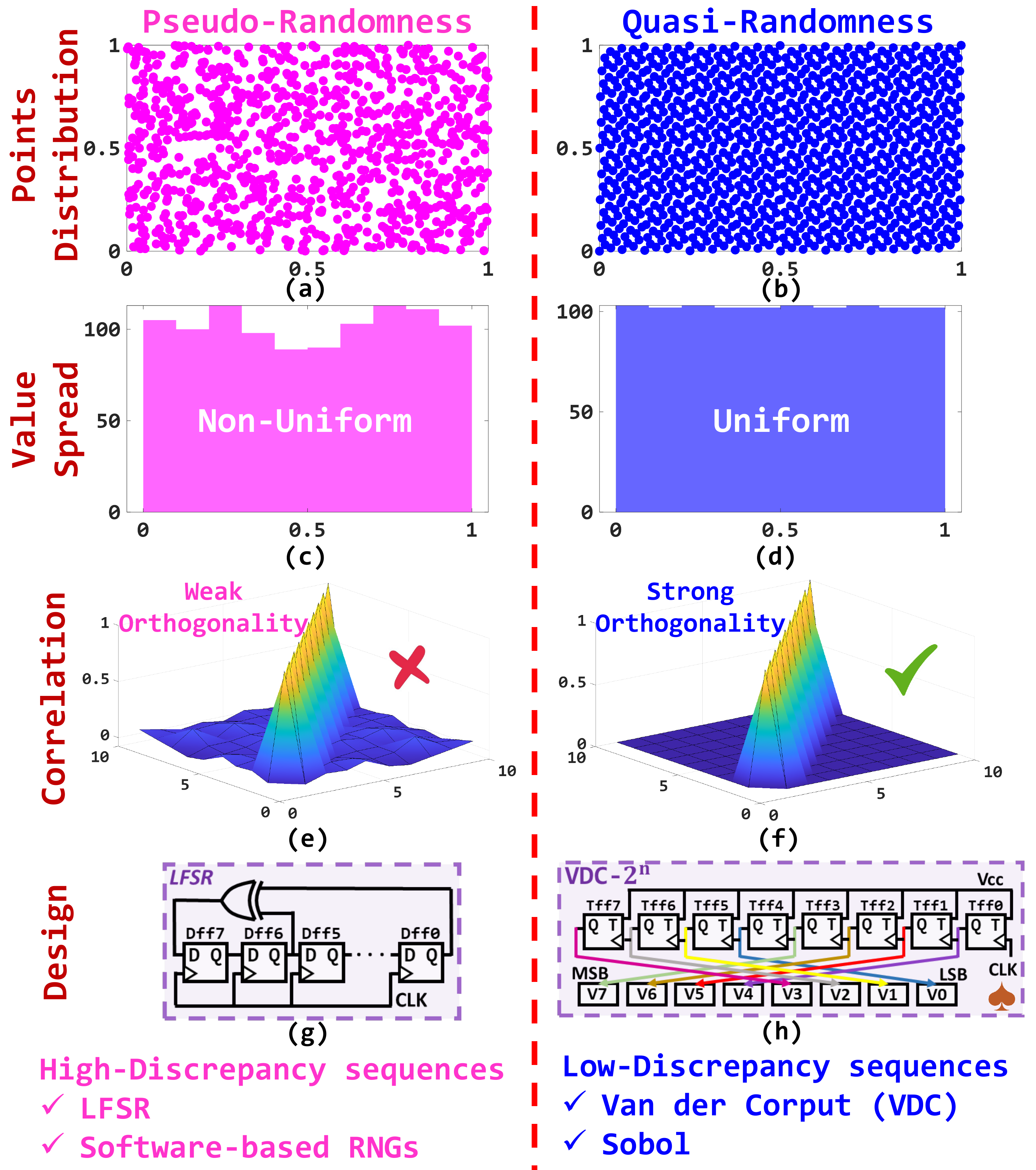}
  \caption{Pseudo-randomness vs. Quasi-randomness in random sequences. (a),(b) \textit{Points Distribution} plots, (c),(d) \textit{Value Spread} plots, (e),(f) \textit{Correlation/Orthogonality} plots, and (g), (h) \textit{Design Implementation}.}
  \label{randomness}
\end{figure}

Ensuring high-quality randomness through 
an appropriate random number generator (RNG) 
is essential in achieving the desired accuracy and hardware efficiency for both SC and HDC systems.
In number theory, pseudo-randomness and quasi-randomness are two well-established concepts. Pseudo-randomness refers to 
sequences or processes that appear statistically random but are generated by deterministic algorithms. Pseudo-random sequences exhibit key randomness characteristics, such as uniform distribution and unpredictability over short intervals, yet they remain reproducible if the generator's initial conditions or seed are known~\cite{pseudo-random_salil}. These sequences are often referred to as \textit{high-discrepancy} sequences, where the discrepancy denotes the deviation of the sequence points from the uniformity~\cite{discrepancy_Chazelle,PILLICHSHAMMER2004301}. Linear Feedback Shift Registers \textit{(LFSRs)} are a well-known source for generating pseudo-random sequences.

Conversely, quasi-random sequences, such as Sobol and Halton sequences, 
offer a more even and uniform distribution. These sequences are generated by a special class of deterministic algorithms designed to fill a space (such as a hypercube for multidimensional sequences) more uniformly than pseudo-randomness sequences. Quasi-random 
sequences possess 
a \textit{low-discrepancy} (LD) property, where lower discrepancy leads to enhanced uniformity. Figs.~\ref{randomness}(a) and (b) illustrate the point distribution for pseudo-random and quasi-random sequences, respectively, highlighting the equal distribution of sequence points in quasi-random sequences. 
Figs.~\ref{randomness}(c) and (d) compare the 
value spread
of points in these 
sequences, with quasi-random sequences demonstrating a more uniform distribution. 

Another important factor that signifies \textit{proper} randomness is the degree of correlation between pairs of $\mathcal{BS}$ or $\mathcal{HV}$ as illustrated in Figs.~\ref{randomness}(e) and (f). The level of correlation plays a critical role in the quality of results. 
For instance, SC \textit{multiplication} using bit-wise \texttt{AND} 
requires independent (or uncorrelated) $\mathcal{BS}$s whereas SC \textit{minimum} using the same bitwise operation 
demands highly correlated input $\mathcal{BS}$s~\cite{6657023}.
In HDC systems also, symbolic data (e.g., pixel positions, letter, signal time stamp, etc.) requires orthogonal (uncorrelated) $\mathcal{HV}$s in 
the encoding stage of the learning model~\cite{Mehran-hdc_survey, 10457041}.


Adopting pseudo-randomness for encoding data into $\mathcal{BS}$s or $\mathcal{HV}$s can lead to model performance degradation and increased hardware costs. Achieving the desired level of accuracy often requires running the model iteratively, which results in increased computational overhead, longer system runtime, 
energy inefficiency, and reduced performance.
Previous studies have shown that pseudo-random sequences perform suboptimally in cascaded circuit architectures, where mid-level correlation among $\mathcal{BS}$s is crucial~\cite{10.1145/3611315.3633265}.
To mitigate these challenges, 
we suggest a novel encoding method based on quasi-randomness, 
aimed at enhancing the overall performance of SC and HDC systems.

\begin{figure*}[t]
  \centering
  \includegraphics[width=\linewidth]{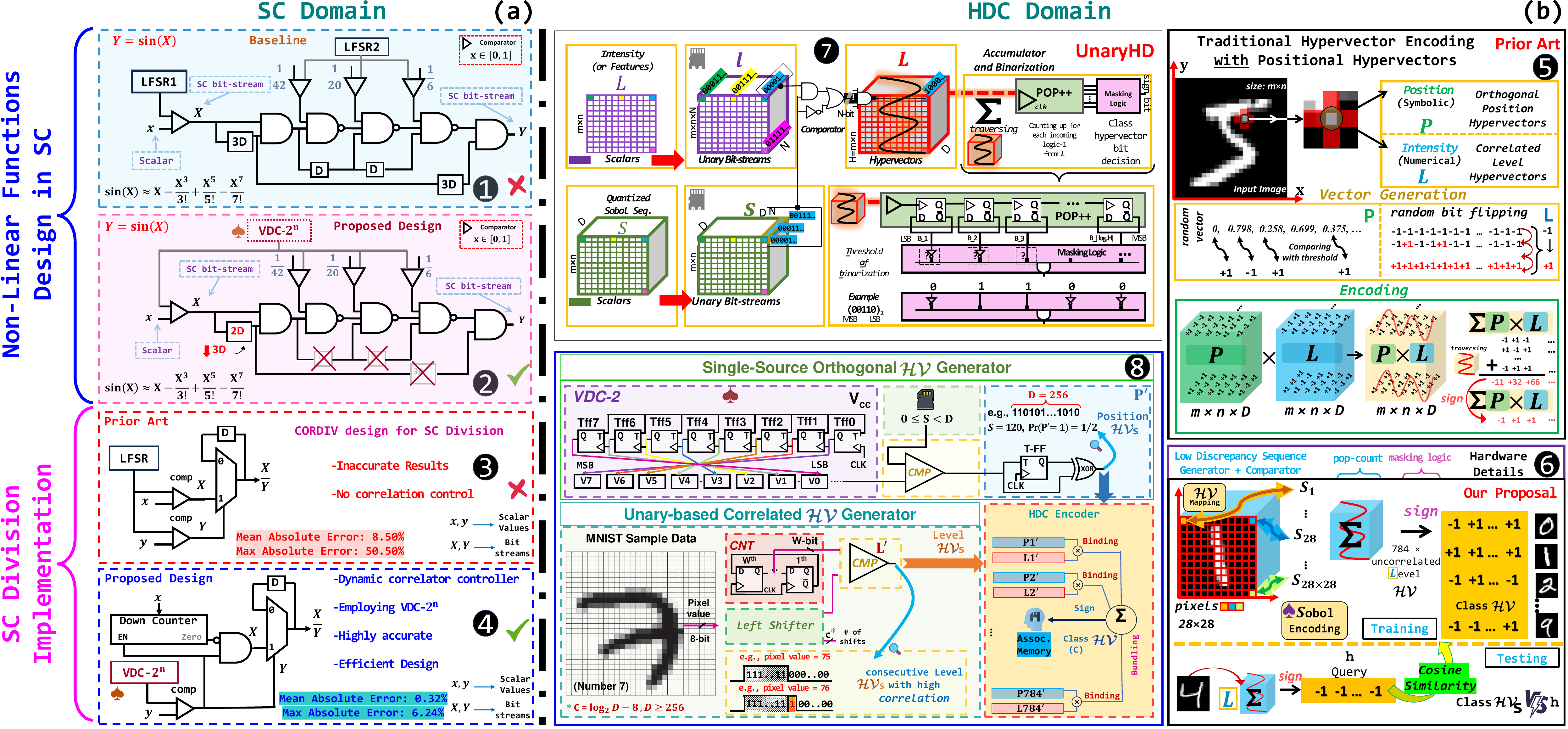}
  \caption{ Applying our \textit{Deterministic} approach to the encoding part of SC and HDC models. (a) Implementing non-linear functions and division operation using the proposed \Design sequence generator in SC and (b) Improving HDC performance by generating high-quality $\mathcal{HV}$s using deterministic sequences. }
  \vspace{-1em}
  \label{designs}
\end{figure*}

\section{Background and Related Work}
In SC, any data value is represented by a sequence of random bits (`0's and `1's)~\cite{10460194,10040569}. 
The probability of `1's appearing in the 
$\mathcal{BS}$ corresponds to the data value. 
A data value $x$ with $n$-bit precision is represented by a $\mathcal{BS}$, 
denoted as $X$, with the probability 
`1's occurring over the entire $\mathcal{BS}$
({\footnotesize $Pr(X=1)/2\textsuperscript{$n$}$}). 
A common method to generate 
$\mathcal{BS}$s involves comparing the given data with  random values from an RNG source. 
While most SOTA works employ LFSRs (Fig.~\ref{randomness}(g)) as RNGs, 
the authors in~\cite{Sobol_TVLSI_2018,Najafi_TVLSI_2019} proposed using Sobol sequences as a deterministic approach for SC, significantly improving the model accuracy.
Another deterministic approach in SC, known as \textit{Unary Computing (UC)}~\cite{8338366,9139000}, utilizes unary $\mathcal{BS}$s where all the `1's are aligned together.
UC is 
free from the random fluctuations of the `0' and `1' bits, which is an important source of error in SC.

Similarly, in HDC, any atomic data unit, called a hypervector ($\mathcal{HV}$), is represented in high dimensionality and comprises elements of `-1's (or `0's) and `1's. In HDC, symbolic data-- such as letters, numbers, sensor data, and temporal and spatial information-- can be represented by distinct and orthogonal $\mathcal{HV}$s. This 
structured information encoding is also known as 
\textit{holographic} representation~\cite{kanerva_holistic}. 
Orthogonality is achieved through randomness
, which generates independent $\mathcal{HV}$s. 
Ideally, an $\mathcal{HV}$ 
consists of an equal number 
of `1's and `0's, with each constituting 50\% of the vector~\cite{10.1145/3538531,9107175,10137195}.

\section{Proposed Framework}
Given the inefficiencies associated with using pseudo-randomness in SC and HDC, 
we propose an efficient, lightweight, and highly accurate deterministic bit-stream generator by utilizing 
Van der Corput (VDC) sequences~\cite{10.1145/3611315.3633265,10473928}. VDC sequences are another example of low discrepancy (LD) sequences that exhibit deterministic yet quasi-random characteristics. In our approach, VDC sequences serve as the primary source of randomness and are ideal candidates for lightweight RNG hardware designs. Generally, VDC sequences are identified by their bases, denoted as $\mathcal{B}$. A VDC-$\mathcal{B}$ sequence number is generated by reversing the digits in the base-$\mathcal{B}$ numeral system, 
resulting in a value within the $[0,1)$ interval. 
Our proposal employs \textit{Powers-of-2} bases for the VDC sequences (\Design). The advantage of using \Design~sequences lies in their simple hardware design and high accuracy in generating bit-streams. 
Any \Design~sequence can be implemented by hardwiring an $n$-bit counter (including \texttt{T} flip flops) (Fig.~\ref{randomness}(h)).
A distinguishing feature of our proposed design is its ability to produce multiple distinct sequences simultaneously through various hardwiring schemes. Another significant attribute that sets it apart from SOTA methods is that our design achieves high accuracy with a single run, whereas pseudo-random methods 
require multiple executions to attain optimal accuracy. 
This feature enhances the overall efficiency and throughput of the system, which is particularly beneficial for 
resource-constrained devices. 

Fig.~\ref{designs}(a) demonstrates an SC implementation of a non-linear function, specifically, 
\textbf{sin(x)}, 
comparing its conventional design (\ding{202})~\cite{Trig-Parhi} with our modified design 
utilizing 
\Design sequences~(\ding{203})~\cite{TriSC,WUC_TriSC}. 
Additionally, We present the implementation of the basic SC division operation (\ding{204})~\cite{7560183} and its enhanced design structure (\ding{205})~\cite{10415201,Mehran-patent_TNANO} 
using the proposed RNG. 
Incorporating \Design~sequences in the design of SC 
operations~\cite{10473928} and trigonometric functions~\cite{TriSC} significantly enhances accuracy while 
reducing the overall hardware costs.

Similarly, we demonstrate that equipping HDC models with such deterministic sequences enhances overall performance while reducing hardware costs.
Fig.~\ref{designs}(b) exhibits the process of applying our 
method to the encoding stage of HDC. 
while the baseline methods (\ding{206}) incorporate and bind (using element-wise multiplication) both \textit{Position} and \textit{Level} $\mathcal{HV}$s for encoding 
images, employing quasi-random 
sequences to generate $\mathcal{HV}$s eliminates the need for \textit{Position} $\mathcal{HV}$s and subsequent multiplication operations (\ding{207})~\cite{10194306,Mehran-patent_HDC,Mehran-patent_HDC_image}. 

As an extension of this approach, we introduce \textit{UnaryHD} architecture (\ding{208}), where \textit{Unary encoding} is applied to HDC models by employing quantized LD sequences for $\mathcal{HV}$ generation~\cite{Mehran-uHD,Mehran-patent_uHD}. This approach simplifies hardware implementation, provides significant cost savings, and contributes to more efficient data encoding in HDC systems. 
To further improve the performance of HDC systems 
we propose 
an end-to-end unary structure. This streamlined design features 
a lightweight, single-source dynamic $\mathcal{HV}$ generator. 
The primary goal of this 
$\mathcal{HV}$ generator design is to achieve optimal randomness in a single iteration, while aligning with the recurrent nature of the random sequence. 
Unlike the baseline
HDC (using LFSR), our proposed design does not employ multiple
random sequences to generate 
\textit{m} different \textit{D}-sized vectors. 
Instead, we generate a single \textit{D}-sized sequence and use it to create different $\mathcal{HV}$s~\cite{Mehran-islped}. 
Another key contribution of this design is 
a lightweight 
hardware used to generate 
\textit{Level} $\mathcal{HV}$s. 
For the first time in the literature, we generate \textit{Level} $\mathcal{HV}$s not randomly but deterministically using our unary generator, eliminating 
the need for randomness. 
Our proposed design 
includes a left shifter module, an up-counter, and a comparator (\ding{209}). 

\begin{table}[t]
\centering
\caption{Accuracy Evaluation of SC \textbf{sin(x)} Designs (Fig.~\ref{designs}(a)\ding{202},\ding{203}).}
\vspace{-0.5em}
\label{acc_sinx}
 \setlength{\tabcolsep}{1.pt}
 \relscale{0.8}

\begin{tabular}{|c|c|c|c|c|c|c|c|c|c|c|} 
\hline
\multirow{2}{*}{\begin{tabular}[c]{@{}c@{}}\\ \textbf{Func.}\end{tabular}} & \multirow{2}{*}{\begin{tabular}[c]{@{}c@{}}\textbf{}\\\textbf{Design}\\ \textbf{Approach}\end{tabular}} & \multirow{2}{*}{\begin{tabular}[c]{@{}c@{}}\\ \textbf{Polyn. } \\\textbf{order}\end{tabular}} & \multirow{2}{*}{\begin{tabular}[c]{@{}c@{}}\textbf{}\\ \textbf{N \textsuperscript{\ding{61}}}\end{tabular}} & \multicolumn{2}{c|}{\textbf{Sequence type}} & \multicolumn{4}{c|}{\textbf{Total \# of Delay Elements}} & \multirow{2}{*}{\begin{tabular}[c]{@{}c@{}}\\ \textbf{MSE} \\ (\texttimes 10\textsuperscript{-4})\end{tabular}} \\ 
\cline{5-10}
 &  &  &  & \begin{tabular}[c]{@{}c@{}} \textbf{BSG}\textsuperscript{\ding{86}} \\ \textbf{Input} \end{tabular} & \begin{tabular}[c]{@{}c@{}}\textbf{BSG} \\\textbf{Coefficients}\end{tabular} & \begin{tabular}[c]{@{}c@{}}\textbf{@}\\ \textbf{Input}\end{tabular} & \begin{tabular}[c]{@{}c@{}}\textbf{@ 1\textsuperscript{st} }\\\textbf{stage}\end{tabular} & \begin{tabular}[c]{@{}c@{}}\textbf{@ 2\textsuperscript{nd} }\\\textbf{stage}\end{tabular} & \begin{tabular}[c]{@{}c@{}}\textbf{@ 3\textsuperscript{rd} }\\\textbf{stage}\end{tabular} &  \\ 
\hline 
\multirow{2}{*}{\begin{tabular}[c]{@{}c@{}} \textbf{sin(x)} \end{tabular}} & \begin{tabular}[c]{@{}c@{}}\textbf{Proposed \ding{203}}\end{tabular} & 7 & \begin{tabular}[c]{@{}c@{}} 1024\\ 512\\ 256 \end{tabular} & \begin{tabular}[c]{@{}c@{}}VDC4 \end{tabular} & \begin{tabular}[c]{@{}c@{}}VDC128,256,512\\VDC128,256,512\\VDC128\end{tabular} & \textbf{2} & \textbf{0} & \textbf{0} & \textbf{0} & \begin{tabular}[c]{@{}c@{}} \textbf{0.523}\\ 0.582\\0.576\end{tabular} \\ 
\cline{2-11}
 &\begin{tabular}[c]{@{}c@{}} \textbf{Baseline \ding{202}} \end{tabular} & 7 & 1024 & LFSR1 & LFSR2 & 3 & 1 & 1 & 3 & 2.256 \\ 
 
\hline

\end{tabular}
\justify{\scriptsize{\ding{86}: $\mathcal{BS}$ Generator $\|$ \ding{61}:$\mathcal{BS}$ length
}
}
\vspace{-1em}
\end{table}


\begin{table}[!t]
\centering
\caption{Hardware Cost Comparison of SC \textbf{sin(x)} Designs (Fig.~\ref{designs}(a)\ding{202},\ding{203}).
}
\vspace{-0.5em}
\label{hw_sinx}
\setlength\dashlinedash{0.2pt}
 \setlength\dashlinegap{1.5pt}
 \setlength\arrayrulewidth{0.3pt}
 \setlength{\tabcolsep}{1.2pt}
\arrayrulecolor{black}
\ADLnullwidehline
\begin{tabular}{|c|c||c|c|c|c|c|c|} 
\hline
\multirow{2}{*}{\begin{tabular}[c]{@{}c@{}}\\ \textbf{Design} \\ \textbf{Approach}\end{tabular}} & \multirow{2}{*}{\begin{tabular}[c]{@{}c@{}}\\ \textbf{N}\end{tabular}} & \multicolumn{6}{c|}{\textbf{sin(x)}}  \\ 
\cline{3-8}
 &  & \begin{tabular}[c]{@{}c@{}}\textbf{Area} \\ ($\mu$\textbf{m\textsuperscript{2}})\end{tabular} & \begin{tabular}[c]{@{}c@{}}\textbf{CPL} \\\textbf{(ns)}\end{tabular} & \begin{tabular}[c]{@{}c@{}}\textbf{Power} \\ ($\mu$\textbf{W})\end{tabular} & \begin{tabular}[c]{@{}c@{}}\textbf{Energy}\\ \textbf{(pJ)}\end{tabular} & \textbf{ADP} & \textbf{EDP} \\ 
\hline
\begin{tabular}[c]{@{}c@{}}\textbf{Proposed} \textbf{\ding{203}}\end{tabular} & \begin{tabular}[c]{@{}c@{}} 1024 \end{tabular} & \begin{tabular}[c]{@{}c@{}} \textbf{554} \end{tabular} & \begin{tabular}[c]{@{}c@{}} 0.44 \end{tabular} & \begin{tabular}[c]{@{}c@{}} \textbf{812.2} \end{tabular} & \begin{tabular}[c]{@{}c@{}} \textbf{365.9} \end{tabular} & \begin{tabular}[c]{@{}c@{}} \textbf{243.7}  \end{tabular} & \begin{tabular}[c]{@{}c@{}} \textbf{160.9} \end{tabular} \\ 
\hline

\begin{tabular}[c]{@{}c@{}}\textbf{Baseline \ding{202}} \end{tabular} & \begin{tabular}[c]{@{}c@{}} 1024 \end{tabular} & \begin{tabular}[c]{@{}c@{}} 801 \end{tabular} & \begin{tabular}[c]{@{}c@{}} 0.42 \end{tabular} & \begin{tabular}[c]{@{}c@{}} 2178.2 \end{tabular} & \begin{tabular}[c]{@{}c@{}} 936.8 \end{tabular} & \begin{tabular}[c]{@{}c@{}} 336.4  \end{tabular} & \begin{tabular}[c]{@{}c@{}} 393.4 \end{tabular}  \\ 
\hline

\end{tabular}
\end{table}

\section{Results and Contributions}

\begin{table}[!t]
\centering
\caption{Accuracy(\%) comparison of the HDC Design in Fig.~\ref{designs}(b)\ding{207} and the Baseline(\ding{206}) \textbf{HDC}.}
\vspace{-0.5em}
\setlength{\tabcolsep}{2.6pt}

\begin{tabular}{|c|c|c|c|c|} 
\hline
\multicolumn{2}{|c|}{\textbf{Method}} & \textbf{\textbf{\textbf{\textbf{\textbf{\textbf{\textbf{\textbf{Minimum}}}}}}}} & \textbf{\textbf{Average}} & \textbf{\textbf{\textbf{\textbf{Maximum}}}} \\ 
\hline
\multirow{2}{*}{\textbf{D=1K}} & \textbf{Baseline} & 70.65 & 79.09 & 84.89 \\ 
\cline{2-5}
 
 & \textbf{Design of \ding{207} with \textit{Sobol}+\Design} & \multicolumn{3}{c|}{\textbf{85.10}} \\

\hline
\multirow{2}{*}{\textbf{D=2K}} & \textbf{\textbf{Baseline}} & 71.81 & 81.29 & 86.96 \\ 
\cline{2-5}

 & \textbf{\textbf{Design of \ding{207} with \textit{Sobol}+\Design}} & \multicolumn{3}{c|}{\textbf{87.12}} \\
 
\hline
\multirow{2}{*}{\textbf{D=8K}} & \textbf{\textbf{\textbf{\textbf{Baseline}}}} & 86.19 & 87.27 & 87.51 \\ 
\cline{2-5}
 
 & \textbf{\textbf{\textbf{\textbf{Design of \ding{207} with \textit{Sobol}+\Design}}}} & \multicolumn{3}{c|}{\textbf{88.68}} \\
 
\hline
\end{tabular}
\justify{\scriptsize{\textit{SOTA HDCs accuracy} (MNIST): \ding{172} \ding{223}\cite{8801933} 75.40\% (w/o retraining) D=2K \linebreak $\|$ 
\ding{173} \ding{223}\cite{9354795} 86.00\% (w/o retraining) D=10K $\|$ \ding{174} \ding{223} \cite{9458526} 88.00\% (w/ retraining) D=10K  $\|$ \linebreak \ding{175} \ding{223}\cite{QuantHD, duan2022braininspired} 87.38\% (w/ retraining) D=10K.
}}
\label{hdc1}
\vspace{-1em}
\end{table}


\begin{table}[!t]
\centering
\caption{Accuracy(\%) comparison of the HDC Design in Fig.~\ref{designs}(b)\ding{208} and the Baseline(\ding{206}) \textbf{HDC}.}
\vspace{-0.5em}
\setlength{\tabcolsep}{1.2pt}
\begin{tabular}{|c|c|c|c|c|c|c|} 
\hline
\multirow{2}{*}{\textbf{Datasets}} & \multicolumn{2}{c|}{\textbf{D=1K}} & \multicolumn{2}{c|}{\textbf{D=2K}} & \multicolumn{2}{c|}{\textbf{D=8K}} \\ 
\cline{2-7}
 & \textbf{\texttt{UnaryHD}} & \textbf{\textit{Baseline}} & \textbf{\texttt{UnaryHD}} & \textbf{\textit{Baseline}} & \textbf{\texttt{UnaryHD}} & \textbf{\textit{Baseline}} \\ 
\hline
\textbf{CIFAR-10} & \textbf{39.29} & 38.21 & \textbf{40.28} & 40.26 & \textbf{41.97} & 41.71 \\ 
\hline
\textbf{Blood MNIST} & \textbf{53.05} & 48.52 & \textbf{55.86} & 51.20 & \textbf{57.88} & 51.82 \\ 
\hline
\textbf{Breast MNIST} & \textbf{68.59} & 68.47 & \textbf{69.23} & 69.11 & \textbf{71.15} & 70.93 \\ 
\hline
\textbf{Fashion MNIST} & \textbf{68.60} & 54.19 & \textbf{70.06} & 69.97 & \textbf{71.37} & 70.87 \\ 
\hline
\textbf{SVHN} & \textbf{60.29} & 60.06 & \textbf{61.73} & 61.24 & \textbf{62.87} & 62.82 \\
\hline
\end{tabular}
\label{hdc2}
\end{table}

To evaluate the effectiveness of  
our proposed 
approach, 
we first applied it to SC designs. 
Tables~\ref{acc_sinx} and \ref{hw_sinx} assess the accuracy and hardware costs for implementing the \textbf{sin($x$)} function. This evaluation highlights the potential of our approach 
for designing SC trigonometric and non-linear functions, which are fundamental components in AI, computer vision, robotics, and communication models~\cite{TriSC, 10637645}. As demonstrated, the proposed design significantly improves the accuracy and reduces energy consumption by 
up to 77\% and 92\%, respectively, compared to the SOTA baseline architecture.

\begin{figure}[!t]
  \centering
  \includegraphics[width=\linewidth]{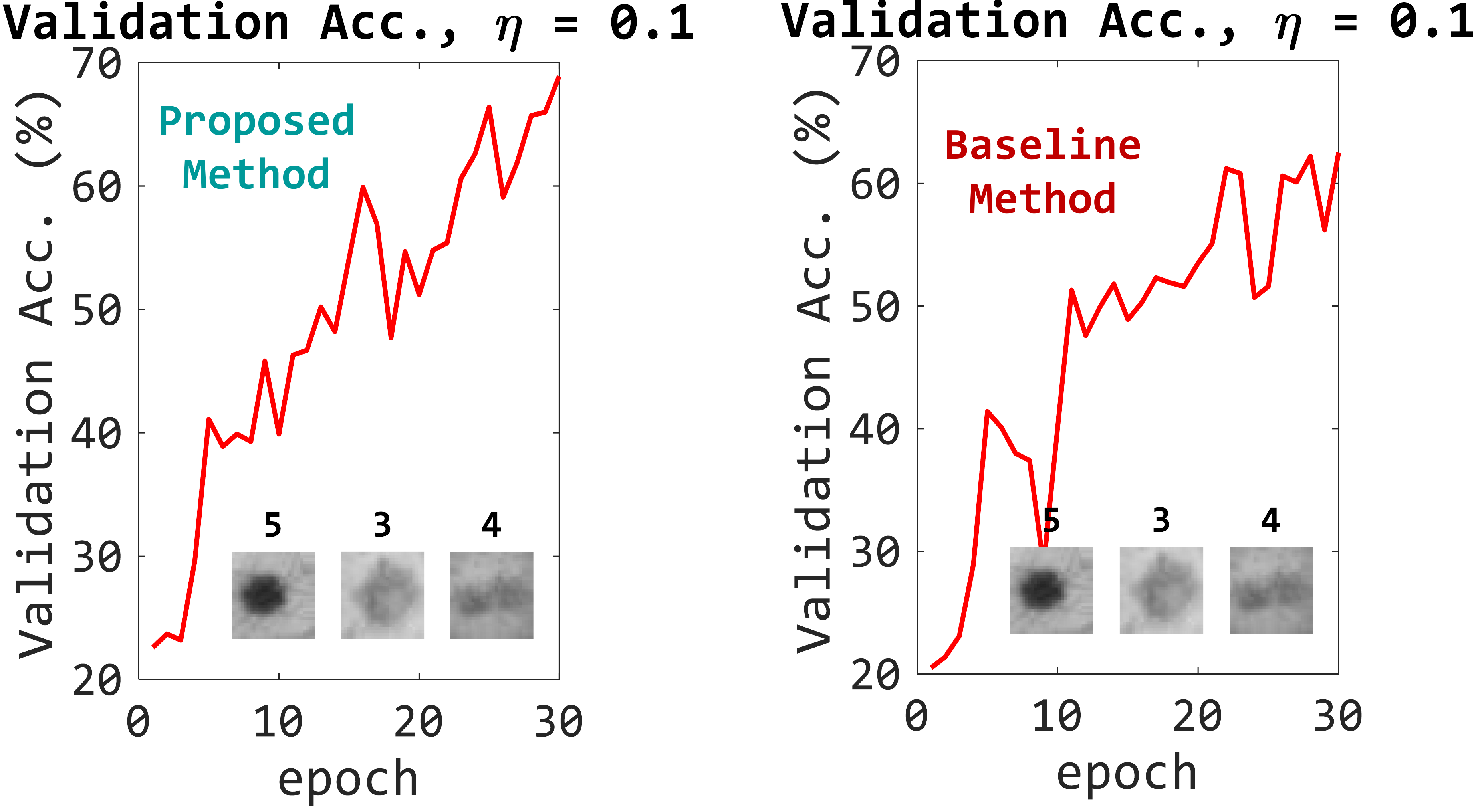}
  \vspace{-1.5em}
  \caption{ Performance evaluation of the end-to-end Unary structure for \textbf{HDC} on the DermaMNIST dataset~\cite{Yang2023} using single-source $\mathcal{HV}$ generator design (Fig.~\ref{designs}(b)\ding{209}), D=1024~\cite{Mehran-islped}. }
  \label{results}
\end{figure}

Employing deterministic sequences to generate high-quality $\mathcal{HV}$s significantly improves the performance of HDC models. Tables~\ref{hdc1} and \ref{hdc2} compare the accuracy  of the designs depicted in Fig.~\ref{designs}(b) (\ding{207}) and (\ding{208}) 
for image classification tasks. 
As shown, the HDC model encoded with deterministic $\mathcal{HV}$s outperforms the baseline model. 
Fig.~\ref{results} showcases the performance 
of the end-to-end unary structure in the HDC model in Fig.~\ref{designs}(b) (\ding{209}) when applied to 
the DermaMNIST dataset~\cite{Yang2023}. Additionally, we incorporated an epoch-based training option, given the increased complexity of this dataset compared to the traditional handwritten digit classification tasks. The results indicate that employing our deterministic solution in HDC encoding facilitates earlier learning progress compared to the baseline model~\cite{10683781}. 
Beyond improved learning dynamics, 
the proposed architecture also offers superior hardware efficiency. 
The proposed $\mathcal{HV}$ generator reduces power consumption by 98\% and improves the energy efficiency by 15\% compared to the baseline method, making it a promising design for dynamic vector generation in resource-constrained edge devices.

\section{Conclusion}
This work presents a novel deterministic sequence generator and explores its significant applicants within the paradigms of Stochastic Computing (SC) and Hyperdimensional Computing (HDC). The key contributions are summarized in four research highlights: 
\ding{192}~Utilizing hardware-friendly quasi-random sequences in SC and HDC systems to generate high-quality bit-streams. 
\ding{193}~Enhancing the model throughput, efficiency, and 
accuracy while simultaneously reducing hardware costs compared to SOTA solutions. \ding{194}~
Introducing a novel, streamlined, and efficient RNG 
for both SC and HDC designs, offering a promising approach for resource-constrained devices. 
\ding{195}~Pioneering the integration of Unary Computing and HDC, resulting in lightweight and energy-efficient HDC systems. 



\bibliographystyle{IEEEtran}
\bibliography{bibliography_,Hassan}

\begin{IEEEbiography}[{\includegraphics[width=1.3in,height=1.25in,clip,keepaspectratio]{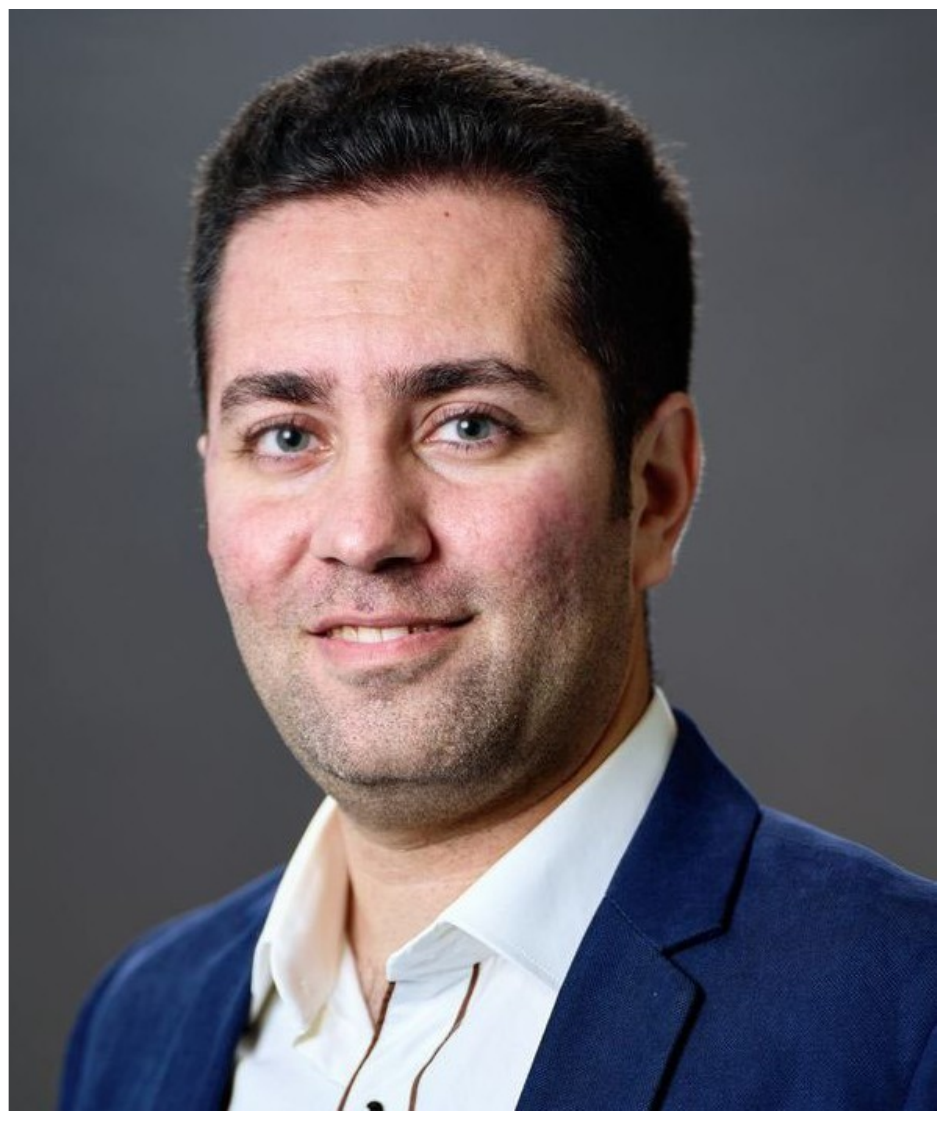}}]{Mehran Shoushtari Moghadam} (S’22) received his B.Sc. degree in Computer Engineering–Hardware and M.Sc. degree in Computer Engineering–Computer Systems Architecture from the University of Isfahan, Iran, in 2010 and 2016, respectively. He graduated as one of the top-ranking students in both programs.
In 2022, he began his Ph.D. studies in Computer Engineering at the School of Computing and Informatics, University of Louisiana at Lafayette, Lafayette, LA, USA. In 2024, he transferred to the Electrical, Computer, and Systems Engineering department at Case Western Reserve University, Cleveland, OH, USA, to continue pursuing his Ph.D. in Computer Engineering.
Mehran became a finalist in the ACM SIGBED Student Research Competition (SRC) at ESWEEK and ICCAD 2024 and was selected as a DAC Young Fellow in DAC 2024. His research interests include emerging and unconventional computing paradigms, such as energy-efficient stochastic computing, real-time and high-accuracy hyperdimensional computing, bit-stream processing, and low-power in/near-sensor computing designs. 
\end{IEEEbiography}


\begin{IEEEbiography}[{\includegraphics[width=1.1in,height=1.25in,clip,keepaspectratio]{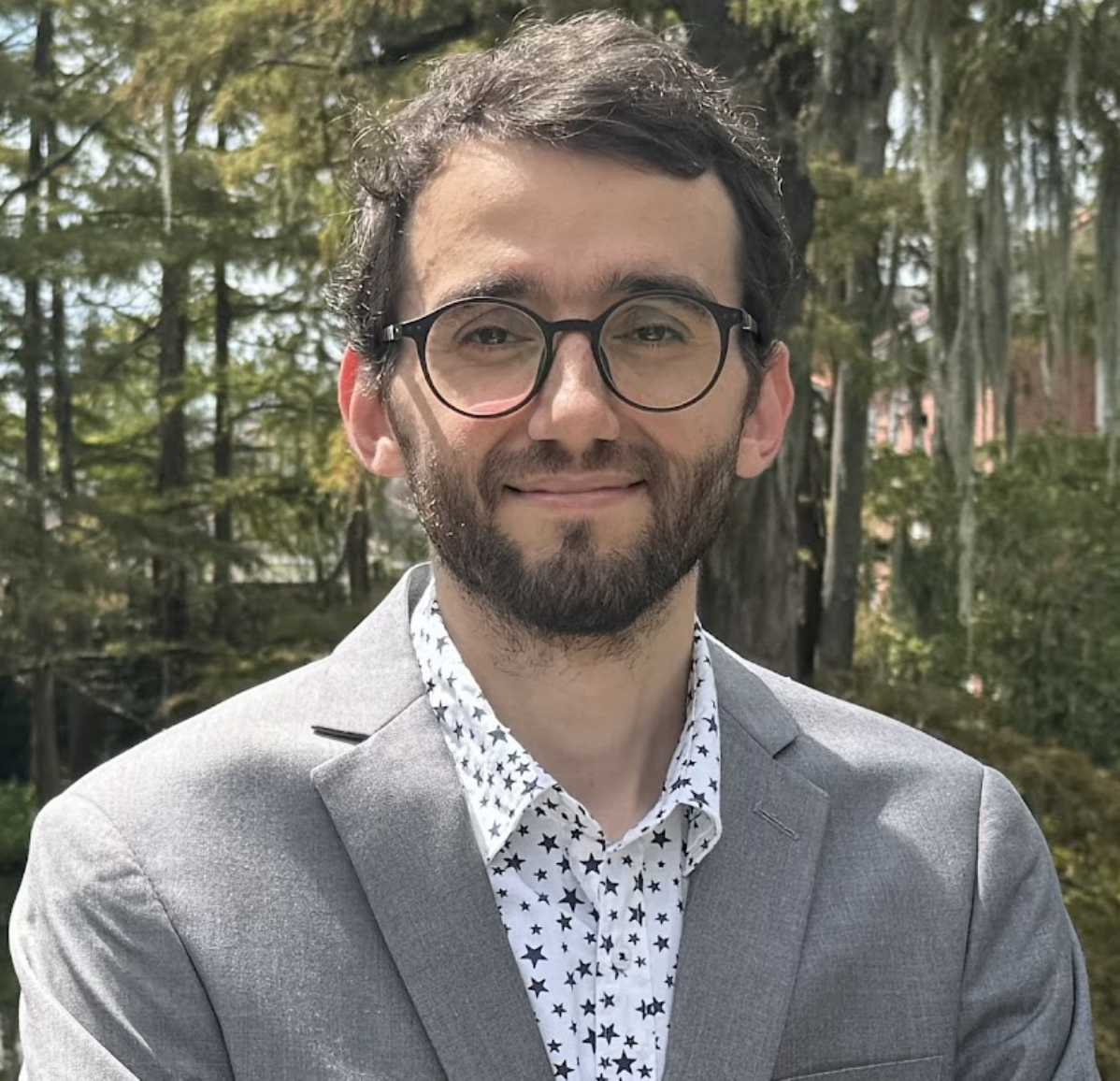}}]{Sercan Aygun} (S’09-M’22-SM'24) received a B.Sc. degree in Electrical \& Electronics Engineering and a double major in Computer Engineering from Eskisehir Osmangazi University, Turkey, in 2013. He completed his M.Sc. degree in Electronics Engineering from Istanbul Technical University in 2015 and a second M.Sc. degree in Computer Engineering from Anadolu University in 2016. Dr. Aygun received his Ph.D. in Electronics Engineering from Istanbul Technical University in 2022. 
Dr. Aygun received the Best Scientific Research Award of the ACM SIGBED Student Research Competition (SRC) ESWEEK 2022, the Best Paper Award at GLSVLSI'23, and the Best Poster Award at GLSVLSI'24. Dr. Aygun's Ph.D. work was recognized with the Best Scientific Application Ph.D. Award by the Turkish Electronic Manufacturers Association and was also ranked first nationwide in the Science and Engineering Ph.D. Thesis Awards by the Turkish Academy of Sciences. He is currently an Assistant Professor at the School of Computing and Informatics, University of Louisiana at Lafayette. He works on tiny machine learning and emerging computing, including stochastic and hyperdimensional computing. \end{IEEEbiography}



\begin{IEEEbiography}[{\includegraphics[width=1.3in,height=1.25in,clip,keepaspectratio]{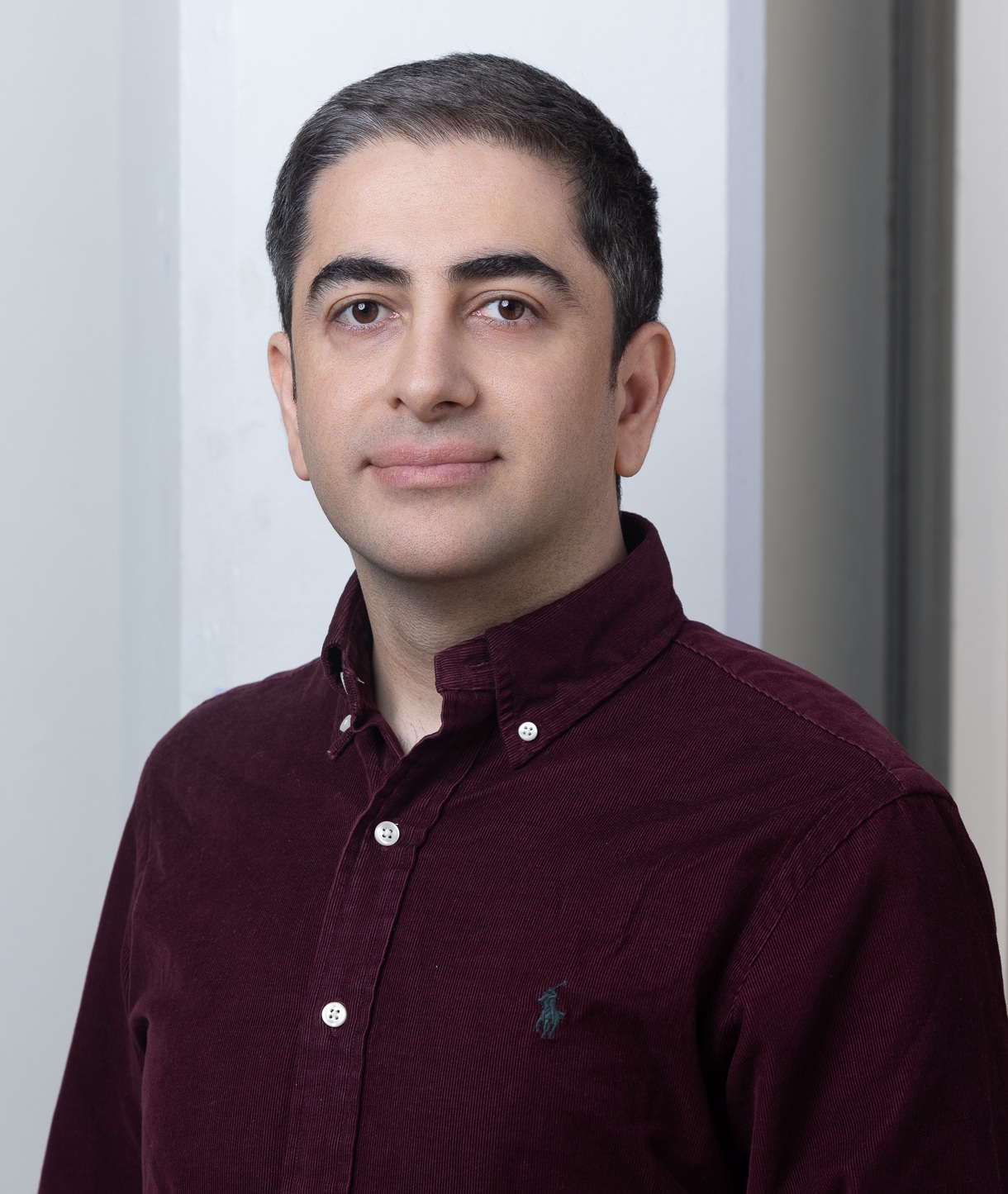}}]{M. Hassan Najafi} (S’15-M’18-SM'23) received the B.Sc. degree in Computer Engineering from the University of Isfahan, Iran, the M.Sc. degree in Computer Architecture from the University of Tehran, Iran, and the Ph.D. degree in Electrical Engineering from the University of Minnesota, Twin Cities, USA, in 2011, 2014, and 2018, respectively. He was an Assistant Professor at the School of Computing and Informatics, University of Louisiana at Lafayette from 2018 to 2024. He is currently an Assistant Professor at the Electrical, Computer, and Systems Engineering Department at Case Western Reserve University. His research interests include stochastic and approximate computing, unary processing, in-memory computing, and hyperdimensional computing. He has authored/co-authored more than 80 peer-reviewed papers and has been granted 6 U.S. patents with more pending. Dr. Najafi received the NSF CAREER Award in 2024, the Best Paper Award at GLSVLSI'23 and ICCD’17, the Best Poster Award at GLSVLSI'24, the 2018 EDAA Outstanding Dissertation Award, and the Doctoral Dissertation Fellowship from the University of Minnesota. Dr. Najafi has been an editor for the IEEE Journal on Emerging and Selected Topics in Circuits and Systems and a Technical Program Committee Member for many EDA conferences. \end{IEEEbiography}

\end{document}